\documentclass[conference]{IEEEtran}
\IEEEoverridecommandlockouts
\usepackage{cite}
\usepackage{amsmath,amssymb,amsfonts}
\usepackage{algorithmic}
\usepackage{graphicx}
\usepackage{textcomp}
\usepackage{xcolor}
\usepackage{booktabs} 
\usepackage[ruled,vlined,linesnumbered]{algorithm2e}
\usepackage{tablefootnote}
\usepackage{subcaption}
\usepackage{multirow}
\usepackage{balance}

\DeclareMathOperator*{\argmax}{argmax}

\newtheorem{problem}{\noindent Problem}
\newtheorem{lm}{Lemma}
\newtheorem{Theorem}{Theorem}
\newtheorem{define}{Definition}

\newcommand{\hide}[1]{}

\newcommand{\aurora}{\textsc{Aurora}}

\newcommand{\hh}[1]{{\small\color{red}{\bf hh: #1}}}

\def\BibTeX{{\rm B\kern-.05em{\sc i\kern-.025em b}\kern-.08em
    T\kern-.1667em\lower.7ex\hbox{E}\kern-.125emX}}
\begin{document}

\title{AURORA: Auditing PageRank on Large Graphs
}

\author{\IEEEauthorblockN{Jian Kang}
\IEEEauthorblockA{\textit{Arizona State University} \\
jkang51@asu.edu} \\
\IEEEauthorblockN{Yinglong Xia}
\IEEEauthorblockA{\textit{Huawei Research America} \\
yinglong.xia@huawei.com} \\
\and
\IEEEauthorblockN{Meijia Wang}
\IEEEauthorblockA{\textit{Arizona State University} \\
mwang164@asu.edu} \\
\IEEEauthorblockN{Wei Fan}
\IEEEauthorblockA{\textit{Tencent} \\
wei.fan@gmail.com} \\
\and
\IEEEauthorblockN{Nan Cao}
\IEEEauthorblockA{\textit{Tongji University} \\
nan.cao@gmail.com} \\
\IEEEauthorblockN{Hanghang Tong}
\IEEEauthorblockA{\textit{Arizona State University} \\
hanghang.tong@asu.edu}
}

\maketitle

\begin{abstract}
Ranking on large-scale graphs plays a fundamental role in many high-impact application domains, ranging from information retrieval, recommender systems, sports team management, biology to neuroscience and many more. PageRank, together with many of its random walk based variants, has become one of the most well-known and widely used algorithms, due to its mathematical elegance and the superior performance across a variety of application domains. Important as it might be, state-of-the-art lacks an intuitive way to {\em explain} the ranking results by PageRank (or its variants), e.g., {\em why} it thinks the returned top-{\em k} webpages are the most important ones in the entire graph; {\em why} it gives a higher rank to actor {\tt John} than actor {\tt Smith} in terms of their relevance w.r.t. a particular movie?

In order to answer these questions, this paper proposes a paradigm shift for PageRank, from identifying {\em which} nodes are most important to understanding {\em why} the ranking algorithm gives a particular ranking result. We formally define the PageRank auditing problem, whose central idea is to identify a set of key graph elements (e.g., edges, nodes, subgraphs) with the highest influence on the ranking results. We formulate it as an optimization problem and propose a family of effective and scalable algorithms (\aurora) to solve it. Our algorithms measure the influence of graph elements and incrementally select influential elements w.r.t. their gradients over the ranking results. We perform extensive empirical evaluations on real-world datasets, which demonstrate that the proposed methods (\aurora) provide intuitive explanations with a linear scalability. 

\end{abstract}


\section{Introduction}\label{sec:intro}
Ranking on graph data is a way to measure node importance and plays a fundamental role in many real-world applications, ranging from information retrieval~\cite{ilprints422}, recommender systems~\cite{gori2007}, social networks~\cite{weng2010}, sports team management~\cite{radicchi2011} to biology~\cite{singh2007} and neuroscience~\cite{crofts2011}. Among others, PageRank~\cite{ilprints422}, together with many of its random walk based variants, is one of the most well-known and widely used ones. Its mathematical elegance lies in that it only requires the topological structure and the associated edge weights as the input. Such a generality makes it applicable to networks\footnote{In this paper, we use `graph' and `network' interchangably.} from many different application domains.

PageRank has a strong ability answering questions like \textit{what} is the most important page in the World Wide Web; \textit{who} is the most influential person in a collaboration network. Despite its superior performance on graph ranking, PageRank lacks intuitive ways to give answers to questions like \textit{why} the top-$k$ returned webpages are the most important ones; \textit{why} actor {\tt John} ranks higher than actor {\tt Smith} in terms of their relevance w.r.t. a particular movie.  \textit{How} the ranking results are derived from the underlying graph structure has largely remained opaque to the end users, who are often not experts in data mining and mathematics.

\begin{figure}[t]

  \begin{subfigure}{0.23\textwidth}
  \includegraphics[width=\textwidth]{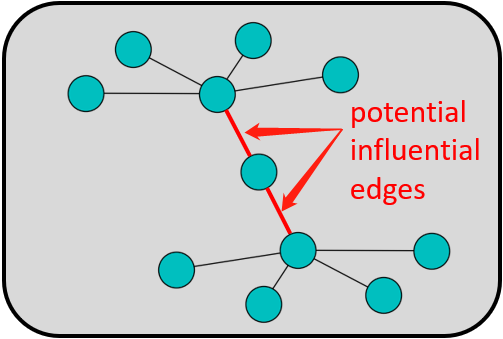}
  \vspace{-6mm}
  \caption{Potential influential edges.}
  \label{fig:exp_edge}
  \end{subfigure}
  \begin{subfigure}{0.23\textwidth}
    \includegraphics[width=\textwidth]{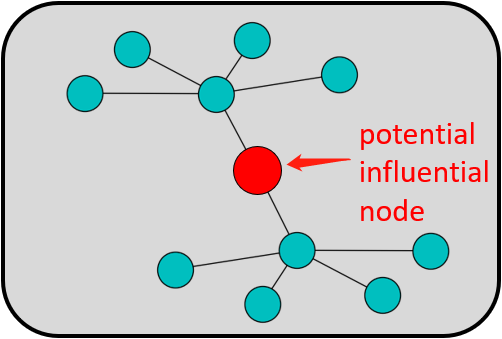}
    \vspace{-6mm}
    \caption{Potential influential node.}
    \label{fig:exp_node}
  \end{subfigure}
  \vspace{-2mm}
  \caption{Example of potential influential edges and node.}
  \vspace{-7mm}
  \label{fig:example}
\end{figure}


To address this challenge, we aim to explain PageRank by finding the influential graph elements (e.g., edges, nodes and subgraphs), which we formulate as {\em PageRank auditing} problem. The key idea is to quantitatively understand how the ranking results would change if we perturb a specific graph element. To be specific, we measure the influence of each graph element by the rate of change in a certain loss function (e.g., $L_p$ norm, etc.) defined over the ranking vector. We believe that auditing graph ranking can benefit many real-world applications. First, it can render the crucial explainability of such ranking algorithms, by identifying valuable information of influential graph elements. Thus, it can help answer questions like why a given node ranks on the top of the ranking list, which link leads to the ranking vector in a certain way. Furthermore, users can use the auditing results to optimize the network topology. In addition, it may help identify the vulnerabilities in the network (e.g. links between two clusters, cutpoints of clusters as shown in Figure~\ref{fig:exp_edge} and Figure~\ref{fig:exp_node}). With the auditing result, users may find several links or nodes that have the greatest influence on the ranking results. It can help users identify if there exist suspicious individuals that manipulate the ranking results by linking heavily with unrelated, off-query topics. Finally, with the knowledge of graph ranking auditing, we may design a more robust ranking algorithm that is hard to be manipulated by users with strategies like linking heavily \cite{Gleich-2015-robustifying}.

The main contributions of the paper are summarized as follows.
\begin{itemize}
  \item {\bf Problem Definition}. We formally define PageRank auditing problem, and formulate it as an optimization problem, whose key idea is to measure the influence of different graph elements as the rate of change in the PageRank results. 
  \item {\bf Algorithms and Analysis}. We propose fast approximation algorithms to solve the PageRank auditing problem.
The algorithms achieve a $(1-1/e)$ approximation ratio with a linear complexity.
  \item {\bf Empirical Evaluations}. We perform extensive experiments on diverse, real-world datasets. The experimental results demonstrate that our proposed methods (a) provide reasonable and intuitive information to find influential graph elements, and (b) scale linearly w.r.t. the graph size. 
\end{itemize}

The rest of the paper is organized as follows. Section~\ref{sec:prob} formally defines the PageRank auditing problem. Section~\ref{sec:algorithm} introduces our proposed algorithms. Then we provide experimental evaluations in Section~\ref{sec:exp}. After reviewing related work in Section~\ref{sec:related}, we conclude the paper in Section~\ref{sec:conclusion}.

\section{Problem Definition}\label{sec:prob}
In this section, we first present a table of symbols (Table 1) that contains the main notations used throughout the paper, and then review PageRank for ranking nodes on graphs, and finally give a formal definition of the PageRank auditing problem.

\begin{table}
  \centering
  \caption{Table of Symbols}
  \label{tab:symbols}
  \begin{tabular}{c|cl}
    \toprule
    Symbols & Definitions \\
    \midrule
    ${\mathbf G} = ({\mathcal V}, {\mathcal E})$ & the input network \\
    $(i, j)$ & edge from node $i$ to node $j$ \\
    ${\mathbf A}$ & adjacency matrix of the input network \\
    ${\mathbf A}(i, j)$ & the element at $i^{\rm th}$ row and $j^{\rm th}$ column \\
    ${\mathbf A}(i, :)$ & $i^{\rm th}$ row of matrix ${\mathbf A}$ \\
    ${\mathbf A}(:, j)$ & $j^{\rm th}$ column of matrix ${\mathbf A}$ \\
    ${\mathbf A}'$ & transpose of the matrix ${\mathbf A}$\\
    ${\mathbf A}^{-1}$ & inverse of the matrix ${\mathbf A}$ \\
    ${\mathbf e}$ & the teleportation vector in PageRank \\
    ${\mathbf r}$ & PageRank of the input network \\
    ${\mathbf r}\left(i\right)$ & ranking score of $i$ \\
    $\textrm{Tr}\left({\mathbf A}\right)$ & Trace of the matrix ${\mathbf A}$ \\
    $\textit{f}\left({\mathbf r}\right)$ & a loss function over ranking vector ${\mathbf r}$ \\
    $n$ & number of nodes in the input network \\
    $m$ & number of edges in the input network \\
    $c$ & damping factor in PageRank \\
    \bottomrule
  \end{tabular}
  \vspace{-4mm}
\end{table}

We use bold upper-case letters for matrices (e.g., ${\mathbf A}$), bold lower-case letters for vectors (e.g., ${\mathbf a}$), calligraphic fonts for sets (e.g., $\mathcal{S}$), and lower-case letters for scalars (e.g., $c$). For matrix indexing conventions, we use the rules similar to Matlab that are shown as follows. We use ${\mathbf A}(i, j)$ to denote the entry of matrix ${\mathbf A}$ at $i^{\rm th}$ row and $j^{\rm th}$ column, ${\mathbf A}(i, :)$ to denote the $i^{\rm th}$ row of matrix ${\mathbf A}$, and ${\mathbf A}(:, j)$ to denote the $j^{\rm th}$ column of matrix ${\mathbf A}$. We use prime to denote the transpose of matrix (i.e., ${\mathbf A}'$ is the transpose of matrix ${\mathbf A}$).


Given a graph ${\mathbf G}$ with $n$ nodes and $m$ edges, PageRank essentially solves the following linear system,
\vspace{-2mm}
\begin{equation}
  {\mathbf r} = c{\mathbf A}{\mathbf r} + (1 - c){\mathbf e}
  \vspace{-2mm}
  \label{eq:random_walk}
\end{equation}
where ${\mathbf e}$ is the teleportation vector with length $n$ and ${\mathbf A}$ is the normalized adjacency matrix of the input graph. In PageRank, ${\mathbf e}$ is chosen as the uniform distribution $\frac{1}{n}{\mathbf 1}$; in personalized PageRank, ${\mathbf e}$ is a biased vector which reflects user's preference (i.e., `personalization') \cite{haveliwala2002topic}; in random walk with restart \cite{tong2006fast}, all the probabilities are concentrated on a single node. A default choice for the normalized adjacency matrix ${\mathbf A}$ is the column-normalized matrix (i.e., the stochastic matrix). A popular alternative choice is the normalized graph Laplacian matrix. In fact, as long as the largest eigenvalue of ${\mathbf A}$ is less than $1/c$, a fix-point solution of the above linear system will converge to  ${\mathbf r} = (1 - c)({\mathbf I} - c{\mathbf A})^{-1}{\mathbf e}$ \cite{li2016quint}. In this paper, we use this general form for the adjacency matrix ${\mathbf A}$. For the ease of description, we also define ${\mathbf Q}=({\mathbf I} - c{\mathbf A})^{-1}$, and ${\mathbf r} = \textrm{pg}({\mathbf A}, {\mathbf e}, c)$ as the resulting PageRank vector with ${\mathbf A}$, ${\mathbf e}$, and $c$ as the corresponding inputs.

Regarding explainable learning and mining techniques, Pang et al. \cite{pang2017} propose a novel notation of influence functions to quantify the impact of each training example on the underlying learning system (e.g., a classifier). The key idea is to trace the model's prediction back to its training data, where the model parameters were derived. In this way, it learns how a perturbation of a single training data will affect the resulting model parameters, and then identifies the training examples that are most responsible for model's predictions. 


Our proposed method to explain the PageRank results is built upon the principle outlined in \cite{pang2017}. To be specific, we aim to find a set of graph elements (e.g., edges, nodes, a subgraph) such that, when we perturb/remove them, the ranking vector will have the greatest change. Formally, we define PageRank auditing problem as follows:
\begin{problem}\label{prob:aurora}
	PageRank Auditing Problem.
\end{problem}

\textbf{Given}: a graph with adjacency matrix ${\mathbf A}$, PageRank vector ${\mathbf r}$, a loss function $f$ over its PageRank vector, user-specific element type (e.g. edges vs. nodes vs. subgraphs), and an integer budget $k$;
\par
\textbf{Find}: a set of $k$ influential graph elements that has the largest impact on the loss function over its PageRank vector $f\left({\mathbf r}\right)$.

\section{Proposed Algorithm}\label{sec:algorithm}
In this section, we propose a family of algorithms (\aurora), to solve PageRank auditing problem (Problem~\ref{prob:aurora}), together with some analysis in terms of effectiveness as well as efficiency. 

\subsection{Formulation}
\par
The intuition behind the proposed methods is to find a set of key graph elements (e.g., edges, nodes, subgraphs) whose perturbation/removal from the graph would affect the PageRank results most. To be specific, let $\mathbf{r} = \textrm{pg}(\mathbf{A}, \mathbf{e}, c)$ be the PageRank vector of the input graph $\mathbf A$, and $\mathbf{r}_{\mathcal{S}} = \textrm{pg}(\mathbf{A}_{\mathcal{S}}, \mathbf{e}, c)$ be the new PageRank vector after removing the graph elements in set $\mathcal{S}$. We formulate the PageRank auditing problem as the following optimization problem.
\begin{equation}\label{eq:optimization}
\begin{aligned}
& \underset{{\mathcal S}}{\text{max}}
& & \Delta{f} = (f({\mathbf r}) - f({\mathbf r}_{{\mathcal S}}))^{2} \\
& \text{s.t.}
& & |{\mathcal S}| = k
\end{aligned}
\end{equation}

\par
In order to solve the above optimization problem, we need to answer two key questions including (Q1) how to quantitatively measure the influence of an individual graph element w.r.t. the objective function; and (Q2) how to collectively find a set of $k$ graph elements with the maximal influence. We first present our proposed solution for Q1 in Subsection \ref{sec:influence}, and then propose three different algorithms for Q2 in Subsections \ref{sec:aurora-e}, \ref{sec:aurora-n} and \ref{sec:aurora-s}, depending on the specific type of graph elements (i.e., edges vs. nodes vs. subgraphs).

\subsection{Measuring Graph Element Influence}\label{sec:influence}
To measure how $f({\mathbf r})$ will change if we perturb/remove a specific graph element, we define its influence as the rate of change in $f({\mathbf r})$.
\begin{define}(Graph Element Influence).\label{def:influence}
  The influence of an edge $(i, j)$ is defined as the derivative of the loss function $f({\mathbf r})$ with respect to the edge, i.e., $\mathbb{I}(i, j) = \frac{df({\mathbf r})}{d{\mathbf A}(i, j)}$. The influence of a node is defined as the aggregation of all inbound and outbound edges that connect to the node., i.e., ${\mathbb I}(i) = \sum\limits_{j = 1, j\neq i}^{n}[{\mathbb I}(i, j)+{\mathbb I}(j, i)] + \sum\limits_{j = 1, j=i}^{n}{\mathbb I}(i, j)$. And the influence of a subgraph is defined as the aggregation of all edges in the subgraph $S$, ${\mathbb I}({S}) = \sum\limits_{i, j\in S}{\mathbb I}(i, j)$.
\end{define}

We can see that the influence for both nodes and subgraphs can be naturally computed based on the edge influence. Therefore, we will focus on how to measure the edge influence. By the property of the derivative of matrices, we first rewrite the influence $\frac{df({\mathbf r})}{d{\mathbf A}(i, j)}$ as 
\begin{equation}
\label{eq:gradient}
  \frac{df({\mathbf r})}{d{\mathbf A}} =
  \begin{cases}
    \frac{\partial f({\mathbf r})}{\partial{\mathbf A}} + (\frac{\partial f({\mathbf r})}{\partial{\mathbf A}})' - \textrm{diag}(\frac{\partial f({\mathbf r})}{\partial{\mathbf A}}), & if\ undirected\\
    \frac{\partial f({\mathbf r})}{\partial{\mathbf A}}, & if\ directed
  \end{cases}
\end{equation}
Directly calculating $\frac{\partial f({\mathbf r})}{\partial{\mathbf A}(i, j)}$ is hard, and we resort to the chain rule:
\begin{equation}
  \label{eq:dr_dAij}
  \frac{\partial f({\mathbf r})}{\partial {\mathbf A}(i, j)} = \frac{\partial f({\mathbf r})}{\partial {\mathbf r}} \frac{\partial {\mathbf r}}{\partial {\mathbf A}(i, j)}
\end{equation}

Next, we present the details on how to solve each partial derivative in Eq.~\eqref{eq:dr_dAij} one by one.

\paragraph{Computing $\frac{\partial f({\mathbf r})}{\partial {\mathbf r}}$}

Here we discuss the choices of $f(\cdot)$ function. In this paper, we choose $f(\cdot)$ to be squared $L_2$ norm for simplicity. However, it is worth mentioning that the proposed methods are applicable to a variety of other loss functions as well. We list some alternative choices and their corresponding derivatives in Table~\ref{tab:f_function}. In the table, \textit{$L_p$ norm} is the most commonly-used vector norm that measures the overall sizes of the vector. Some commonly-used $L_p$ norm includes $L_1$ norm and $L_2$  norm (also known as the Euclidean norm); \textit{soft maximum} is used to approximate the maximum value of elements in the vector; \textit{energy norm} is a measurement often used in system and control theory to measure the internal energy of vector. 
\begin{table}
  \caption{Choices of $f(\cdot)$ functions and their derivatives}
  \label{tab:f_function}
  \begin{tabular}{c|c|c}
    \toprule
    Descriptions & Functions & Derivatives \\
    \midrule
    $L_p$ norm & $f({\mathbf r}) = ||{\mathbf r}||_{p}$ & $\frac{\partial f}{\partial{\mathbf r}} = \frac{{\mathbf r}\circ|{\mathbf r}|^{p-2}}{||{\mathbf r}||_{p}^{p-1}}$ \\
    Soft maximum & $f({\mathbf r}) = log(\sum\limits_{i=1}^{n}exp({\mathbf r}(i)))$ & $\frac{\partial f}{\partial{\mathbf r}} = [\frac{exp({\mathbf r}(i))}{\sum\limits_{i=1}^{n}exp({\mathbf r}(i)}]$ \\
    Energy norm & $f({\mathbf r}) = {\mathbf r}'{\mathbf M}{\mathbf r}$ & $\frac{\partial f}{\partial{\mathbf r}} = ({\mathbf M}+{\mathbf M}'){\mathbf r}$ \\
    \bottomrule
    \hline
    \multicolumn{3}{l}{(${\mathbf M}$ in Energy Norm is a Hermitian positive definite matrix.)}
  \end{tabular}
  \vspace{-6mm}
\end{table}

\hide{
\paragraph{Computing $\frac{\partial {\mathbf r}}{\partial {\mathbf Q}}$} 
Recall that ${\mathbf Q} = ({\mathbf I}-c{\mathbf A})^{-1}$, and the closed-form PageRank solution can be written as ${\mathbf r}=(1-c){\mathbf Q}{\mathbf e}$. Then we have that \hh{double check this -- intuititvely, this should be a $n \times n \times n$ tensor?}
\vspace{-2mm}
\begin{equation}
	\frac{\partial {\mathbf r}}{\partial {\mathbf Q}} = (1 - c){\mathbf e}'
    \vspace{-1mm}
\end{equation}

\paragraph{3. Computing $\frac{\partial {\mathbf Q}}{\partial {\mathbf A}(i, j)}$} Note that ${\mathbf Q} = ({\mathbf I}-c{\mathbf A})^{-1}$, and ${\mathbf A}$ is the adjacency matrix of the input graph. By the property of the derivative of matrix inverse, we have that
\begin{equation}
	\frac{\partial {\mathbf Q}}{\partial {\mathbf A}(i, j)} =
    \begin{cases}
    c{\mathbf Q}(:, i){\mathbf Q}(j, :) + c{\mathbf Q}(:, j){\mathbf Q}(i, :), & if\ undirected \\
    c{\mathbf Q}(:, i){\mathbf Q}(j, :), & if\ directed
    \end{cases}
\end{equation}
}

\paragraph{2. Computing $\frac{\partial {\mathbf r}}{\partial {\mathbf A}(i, j)}$}
Recall that ${\mathbf A}$ is the adjacency matrix of the input graph, and the PageRank solution can be written as solving the linear system in Eq.~\eqref{eq:random_walk}. Then we have that 

\begin{equation}\label{eq:partial_eq}
\frac{\partial {\mathbf r}}{\partial {\mathbf A}(i, j)} = c \frac{\partial \mathbf{A}}{\partial {\mathbf A}(i, j)}{\mathbf r} + c\mathbf{A}\frac{\partial {\mathbf r}}{\partial{\mathbf A}(i, j)}
\end{equation}
Move the second term in Eq.~\eqref{eq:partial_eq} to the left, we have that
\begin{equation}
({\mathbf I}-c{\mathbf A})\frac{\partial {\mathbf r}}{\partial {\mathbf A}(i, j)} = c \frac{\partial \mathbf{A}}{\partial {\mathbf A}(i, j)}{\mathbf r}
\end{equation}
\begin{equation}\label{eq:r_gradient}
\frac{\partial {\mathbf r}}{\partial {\mathbf A}(i, j)} = c({\mathbf I}-c{\mathbf A})^{-1}\frac{\partial {\mathbf A}}{\partial {\mathbf A}(i, j)}{\mathbf r}
\end{equation}
By the property of the first order derivative of matrix, we have that 
\begin{equation}
	\frac{\partial {\mathbf A}}{\partial {\mathbf A}(i, j)} =
    \begin{cases}
    {\mathbf S}^{ij} + {\mathbf S}^{ji}, & if\ undirected \\
    {\mathbf S}^{ij}, & if\ directed
    \end{cases}
\end{equation}
where ${\mathbf S}^{ij}$ is the single-entry matrix with $1$ at $i^{\rm th}$ row and $j^{\rm th}$ column and $0$ elsewhere. Recall that ${\mathbf Q} = ({\mathbf I}-c{\mathbf A})^{-1}$, we can re-write Eq.~\eqref{eq:r_gradient} as the following equation.
\begin{equation}
\frac{\partial {\mathbf r}}{\partial {\mathbf A}(i, j)} = c {\mathbf r}(j) \mathbf{Q}(:, i)
\end{equation}
Combine everything together, we get the closed-form solution for calculating the influence of an edge $(i, j)$ as follows:
\begin{equation}
\label{edge:closed_form}
  \frac{\partial f({\mathbf r})}{\partial {\mathbf A}(i, j)} = 2c{\mathbf r}(j)\textrm{Tr}({\mathbf r}'{\mathbf Q}(:, i))
\end{equation}
Following Eq.~\eqref{edge:closed_form}, 
we get the matrix of gradients for all edges as
\begin{equation}
	\label{eq:partial_gradient}
	\frac{\partial f({\mathbf r})}{\partial{\mathbf A}} = 2c(1-c){\mathbf Q}'{\mathbf r}{\mathbf e}'{\mathbf Q}'
\end{equation}

Two major computational challenges in calculating $\frac{\partial f({\mathbf r})}{\partial{\mathbf A}}$ lie in (1) calculating ${\mathbf Q}$ with $O(n^{3})$ time complexity and (2) $O(n^{2})$ space complexity to save the matrix of gradients. We address both challenges by exploring the low-rank structure of $\frac{\partial f({\mathbf r})}{\partial{\mathbf A}}$. From Eq.~\eqref{eq:partial_gradient}, we can show that it can be re-written as the following low-rank form
\begin{equation}
  \label{eq:low_rank}
  \frac{\partial f({\mathbf r})}{\partial{\mathbf A}} = 2c(1-c){\mathbf Q}'{\mathbf r}{\mathbf e}'{\mathbf Q}' = 2c({\mathbf Q}'{\mathbf r}){\mathbf r}'
\end{equation}
where ${\mathbf Q}'{\mathbf r}$ is an $n\times 1$ vector and ${\mathbf r}$ is an $n\times 1$ PageRank vector. Since ${\mathbf r} = (1 - c){\mathbf Q}{\mathbf e}$, we have that ${\mathbf Q}'{\mathbf r}$ is a personalized PageRank on the reverse of the input graph with ${\mathbf r}$ as teleportation vector with a constant scaling. With this in mind, we do not need to calculate ${\mathbf Q}$ explicitly, or to save the entire matrix directly. Instead, we can use power method to calculate ${\mathbf r}$ and ${\mathbf Q}'{\mathbf r}$, each with $O(m)$ time and save these two vectors with $O(n)$ space. To extract the element of $\frac{\partial f({\mathbf r})}{\partial{\mathbf A}}$ at the $i^\textrm{th}$ row and the $j^\textrm{th}$ column, we simply calculate the product of the $i^\textrm{th}$ element in ${\mathbf Q}'{\mathbf r}$ and the $j^\textrm{th}$ element in ${\mathbf r}$, and scale it by $2c$, which takes $O(1)$ time.


\subsection{Auditing by Edges: \aurora-E}\label{sec:aurora-e}
Due to its combinatorial nature, straight-forward methods for solving the optimization problem in Eq.~\eqref{eq:optimization} are not feasible. The key behind the proposed family of algorithms is due to the diminishing returns property of Problem~\ref{prob:aurora}, which is summarized in Theorem~\ref{theorem:submodular}

\begin{Theorem}
\label{theorem:submodular}
  (Diminishing Returns Property of Problem 1). For {\em any} loss function listed in Table 2, and for {\em any} set of graph elements ${\mathcal S}$, which could be either a set of edges, nodes or subgraphs, in the given graph, its influence measure ${\mathbb I({\mathcal S})}$ defined in Definition 1 is (a) normalized; (b) monotonically non-decreasing; (c) submodular, where ${\mathcal S}\subseteq {\mathcal E}$.
\end{Theorem}
\begin{IEEEproof}
We only prove the diminishing returns property in the edge case. The proofs for nodes and subgraphs are similar and thus is omitted due to the space limitation.

  Let ${\mathbb I}({\mathcal S}) = \sum\limits_{(i, j)\in {\mathcal S}}{\mathbb I}(i, j)$. It is trivial that if there is no edge selected, the influence is 0. Thus it is normalized.
  \par
  Let ${\mathcal I}, {\mathcal J}, {\mathcal K}$ be three sets and ${\mathcal I}\subseteq {\mathcal J}$. We further define three sets ${\mathcal S}, {\mathcal T}, {\mathcal K}$ as follows: ${\mathcal S} = {\mathcal I}\cup {\mathcal K}$, ${\mathcal T} = {\mathcal J}\cup {\mathcal K}$ and ${\mathcal R} = {\mathcal J}\setminus {\mathcal I}$, then we have
  \begin{displaymath}
    \begin{split}
      {\mathbb I}({\mathcal J}) - {\mathbb I}({\mathcal I}) & = \sum\limits_{(i, j)\in {\mathcal J}}{\mathbb I}(i, j) - \sum\limits_{(i, j)\in {\mathcal I}}{\mathbb I}(i, j) \\
      & = \sum\limits_{(i, j)\in {\mathcal J}\setminus {\mathcal I}}{\mathbb I}(i, j) \\
      & = \sum\limits_{(i, j)\in {\mathcal R}}{\mathbb I}(i, j) \\
      & \geq 0
    \end{split}
  \end{displaymath}
  which proves that ${\mathbb I}({\mathcal S})$ is monotonically non-decreasing.

Finally, we prove that it is submodular. Define ${\mathcal P} = {\mathcal T}\setminus {\mathcal S}$.  We have that ${\mathcal P}=({\mathcal J}\cup {\mathcal K})\setminus({\mathcal I}\cup {\mathcal K})={\mathcal R}\setminus({\mathcal R}\cap {\mathcal K})\subseteq {\mathcal R}={\mathcal J}\setminus {\mathcal I}$. Then we have
  \begin{displaymath}
    {\mathbb I}({\mathcal T}) - {\mathbb I}({\mathcal S}) = \sum\limits_{(i, j)\in {\mathcal P}}{\mathbb I}(i, j)\leq {\mathbb I}({\mathcal J}) - {\mathbb I}({\mathcal I})
  \end{displaymath}
  which proves the submodularity of the edge influence.
\end{IEEEproof}

The diminishing returns property naturally leads to a greedy algorithm to obtain a near-optimal solution for solving Problem~\ref{prob:aurora}. We first present the algorithm for auditing by edges in this subsection. The algorithms for auditing by nodes and by subgraphs will be presented in Subsections \ref{sec:aurora-n} and \ref{sec:aurora-s}, respectively.


With the diminishing returns property, we propose \aurora-E (Algorithm~\ref{alg:aurora_e}) algorithm to find top-{\em k} influential edges. The key idea of \aurora-E is to select one edge and update the gradient matrix at each of the $k$ iterations.

\begin{algorithm}
 \SetKwInOut{Input}{Input}
 \SetKwInOut{Output}{Output}
 \Input{ The adjacency matrix ${\mathbf A}$, integer budget $k$}
 \Output{ A set of $k$ edges $\mathcal{S}$ with the highest influence}
 initialize $\mathcal{S} = \emptyset$\;
 initialize $c$ (e.g., $c =  1/{2 \max \textrm{eigenvalue}({\mathbf A})}$)\;
 calculate PageRank ${\mathbf r}=\textrm{pg}({\mathbf A}, {\mathbf e}, c)$\;
 calculate partial gradients $\frac{\partial f(\textbf{r})}{\partial\textbf{A}}$ by Eq.~\eqref{eq:low_rank}\;
 calculate gradients $\frac{df(\textbf{r})}{d\textbf{A}}$ by Eq.~\eqref{eq:gradient}\;
 \While{$|\mathcal{S}|\ne k$ }{
  find $(i, j) = \underset{(i, j)}\argmax\ {\mathbb I}(i, j)$ with Eq.~\eqref{eq:gradient}\;
  add edge $(i, j)$ to $\mathcal{S}$\;
  remove $(i, j)$, and remove $(j, i)$ if undirected\;
  re-calculate ${\mathbf r}$, $\frac{\partial f({\mathbf r})}{\partial{\mathbf A}}$ by Eq.~\eqref{eq:low_rank}, and $\frac{df({\mathbf r})}{d{\mathbf A}}$ by Eq.~\eqref{eq:gradient}\;
 }
 \Return $\mathcal{S}$\;
 \caption{\aurora-E}
 \label{alg:aurora_e}
\end{algorithm}

The effectiveness and efficiency of the proposed \aurora-E are summarized in Lemma~\ref{lemma:approximation} and Lemma~\ref{lemma:complexity_edge}, respectively. We can see that \aurora-E finds a $(1-1/e)$ near-optimal solution with a linear complexity.

\begin{lm}
\label{lemma:approximation}
  (Approximation Ratio of \aurora-E).
  Let ${\mathcal S}_{k}=\newline\{s_{1}, s_{2}, ..., s_{k}\}$ represents the set formed by \aurora-E, $\mathcal{O}$ is the optimal solution of Problem~\ref{prob:aurora}, ${\mathbb I}({\mathcal S})$ is the influence defined in Definition~\ref{def:influence}.
  \vspace{-2mm}
  \begin{displaymath}
    {\mathbb I}({\mathcal S}_{k})\geq (1-1/e){\mathbb I}({\mathcal O})
    \vspace{-2mm}
  \end{displaymath}
\end{lm}
\begin{IEEEproof}
  Omitted for space.
  \hide{By diminishing returns property, $\forall\ i\leq k$, we have
  \begin{displaymath}
  \begin{split}
  f(OPT) & \leq f(OPT\cup {\mathcal S}_{i}) \\
  & = f({\mathcal S}_{i}) + \sum\limits_{s\in OPT}\Delta(s|{\mathcal S}_{i}\cup (OPT\setminus\{s\})) \\
  & \leq f({\mathcal S}_{i}) +\sum\limits_{s\in OPT}\Delta(s|{\mathcal S}_{i}) \\
  & \leq f({\mathcal S}_{i}) +k\Delta(s_{max}|{\mathcal S}_{k}) \\
  \end{split}
  \end{displaymath}
  where $s_{max}=\argmax_{s\in {\mathcal V}\setminus {\mathcal S}_{i}}\Delta(s|{\mathcal S}_{i})$. Then we have
  \begin{displaymath}
    \Delta(s_{max}|{\mathcal S}_{k}) = f({\mathcal S}_{i+1}) - f({\mathcal S}_{i}) \geq \frac{1}{k}(f(OPT) - f({\mathcal S}_{i}))
  \end{displaymath}
  After rearranging the terms, we have
  \begin{gather*}
    f(OPT) - f({\mathcal S}_{i+1}) \leq (1-\frac{1}{k})(f(OPT) - f({\mathcal S}_{i})) \\
    f(OPT) - f({\mathcal S}_{i}) \leq (1-\frac{1}{k})(f(OPT) - f({\mathcal S}_{i-1})) \\
    ... \\
    f(OPT) - f({\mathcal S}_{1}) \leq (1-\frac{1}{k})(f(OPT) - f({\mathcal S}_{0}))
  \end{gather*}
  Thus, recursively apply the inequality, we have
  \begin{displaymath}
  \begin{split}
    f(OPT) - f({\mathcal S}_{k}) & \leq (1-\frac{1}{k})^{k}(f(OPT) - f({\mathcal S}_{0})) \\
   & = (1- \frac{1}{k})^{k}f(OPT) \\
   & \leq \frac{1}{e}f(OPT)
  \end{split}
  \end{displaymath}
  Thus we have $(1 - 1/e)f(OPT) \leq f({\mathcal S}_{i+1})$.
  }
\end{IEEEproof}

\begin{lm}
\label{lemma:complexity_edge}
  (Time and Space Complexities of \aurora-E). Algorithm 1 is $O(mk)$ in time and $O(m+n)$ in space, where $m$ and $n$ are the numbers of edges and nodes in the input graph; and $k$ is the budget.
\end{lm}
\begin{IEEEproof}
It takes $O(m)$ time complexity to calculate $\textbf{r}$ and $\frac{\partial f({\mathbf r})}{\partial {\mathbf A}}$ by applying power iterations. In the while-loop, we find the edge with the greatest influence by traversing all edges, which takes $O(m)$ time. Time spent to re-calculate ${\mathbf r}$ and $\frac{\partial f({\mathbf r})}{\partial {\mathbf A}}$ remains the same as $O(m)$. Since the body inside the loop will run $k$ times, the overall time complexity is $O(mk)$. In Algorithm 1, it takes $O(m)$ space to save the sparse adjacency matrix ${\mathbf A}$ and $O(n)$ space to save the PageRank vector ${\mathbf r}$ and column vector $\mathbf{Q}'\mathbf{r}$ in Eq.~\eqref{eq:low_rank}. Therefore it has $O(m+n)$ space complexity.
\end{IEEEproof}

\subsection{Auditing by Nodes: \aurora-N}\label{sec:aurora-n}
By Theorem~\ref{theorem:submodular}, the influence of nodes also enjoys the diminishing returns property. Following this, we propose a greedy algorithm \aurora-N (Algorithm~\ref{alg:aurora_n}) to find a set of $k$ influential nodes with $(1-1/e)$ approximation ratio with a linear complexity. The efficiency of the proposed \aurora-N is summarized in Lemma~\ref{lemma:complexity_node}.
\begin{algorithm}
  \SetKwInOut{Input}{Input}
  \SetKwInOut{Output}{Output}
  \Input{The adjacency matrix $\textbf{A}$, integer budget $k$}
  \Output{ A set of $k$ nodes $\mathcal{S}$ with highest influence}
  initialize $\mathcal{S} = \emptyset$\;
  initialize $c$ (e.g., $c =  1/{2 \max \textrm{eigenvalue}({\mathbf A})}$)\;
  calculate PageRank ${\mathbf r}=\textrm{pg}({\mathbf A}, {\mathbf e}, c)$\;
  calculate partial gradients $\frac{\partial f({\mathbf r})}{\partial {\mathbf A}}$ by Eq.~\eqref{eq:low_rank}\;
  calculate gradients $\frac{df({\mathbf r})}{d{\mathbf A}}$ by Eq.~\eqref{eq:gradient}\;
  \While{$|\mathcal{S}|\ne k$ }{
    find $v_{i} = \underset{i}\argmax\ {\mathbb I}(i)$\;
    add $v_{i}$ to $\mathcal{S}$\;
    remove all inbound and outbound edges of $v_{i}$\;
    re-calculate ${\mathbf r}$, $\frac{\partial f({\mathbf r})}{\partial {\mathbf A}}$ by Eq.~\eqref{eq:low_rank}, and $\frac{df({\mathbf r})}{d{\mathbf A}}$ by Eq.~\eqref{eq:gradient}\;
  }
  \Return $\mathcal{S}$\;
  \caption{\aurora-N}
  \label{alg:aurora_n}
\end{algorithm}


\begin{lm}
\label{lemma:complexity_node}
  (Time and Space Complexities of \aurora-N). Algorithm 2 is $O(mk)$ in time and $O(m+n)$ in space, where $m$ and $n$ are the numbers of edges and nodes in the input graph; and $k$ is the budget.
\end{lm}
\begin{IEEEproof}
\hide{ 
It takes $O(m)$ time complexity to calculate ${\mathbf r}$ and $\frac{\partial f({\mathbf r})}{\partial {\mathbf A}}$ by applying power iterations. In the while-loop, we calculate the influence of nodes and find the node with the greatest influence by traversing all edges, which takes $O(m)$ time. Time spent to re-calculate ${\mathbf r}$ and $\frac{\partial f({\mathbf r})}{\partial {\mathbf A}}$ remains the same as $O(m)$. Since the body inside loop will run $k$ times, the overall time complexity is $O(mk)$. In Algorithm 2, it takes $O(m)$ space to save the sparse adjacency matrix ${\mathbf A}$ and $O(n)$ space to save the PageRank vector ${\mathbf r}$ and column vector $\mathbf{Q}'\mathbf{r}$ in Eq.~\eqref{eq:low_rank}. Therefore it has $O(m+n)$ space complexity.
}
Omitted for space.
\end{IEEEproof}

\subsection{Auditing by Subgraphs: \aurora-S}\label{sec:aurora-s}
Here, we discuss how to select an influential subgraph with $k$ nodes and we focus on the vertex-induced subgraph. 
With the diminishing returns property (Theorem~\ref{theorem:submodular}) in mind, we propose \aurora-S (Algorithm~\ref{alg:aurora_s}) to greedily identify the influential subgraph with $(1-1/e)$ approximation ratio with a linear complexity. The efficiency of the proposed \aurora-S is summarized in Lemma~\ref{lemma:complexity_subgraph}.

\begin{algorithm}
 \SetKwInOut{Input}{Input}
 \SetKwInOut{Output}{Output}
 \Input{ The adjacency matrix ${\mathbf A}$, output size $k$}
 \Output{ A vertex-induced subgraph of $k$ nodes $\mathcal{S}$ with highest influence}
 initialize $\mathcal{S} = \emptyset$\;
 initialize $c$ (e.g., $c =  1/{2 \max \textrm{eigenvalue}({\mathbf A})}$)\;
 calculate PageRank ${\mathbf r} = \textrm{pg}({\mathbf A}, {\mathbf e}, c)$\;
 calculate partial gradients $\frac{\partial f(\textbf{r})}{\partial\textbf{A}}$ by Eq.~\eqref{eq:low_rank}\;
 calculate gradients $\frac{df(\textbf{r})}{d\textbf{A}}$ by Eq.~\eqref{eq:gradient}\;
 \While{$|\mathcal{S}|\ne k$ }{
  find $(i, j) = \underset{(i, j)}\argmax\ {\mathbb I}(i, j)$\;
  \eIf{$|\mathcal{S}|+2\leq k$}{
    add $v_{i}$ and $v_{j}$ to $\mathcal{S}$\;
    }{
    find the endpoint $v$ with higher gradient\;
    \eIf{$v \not\in\mathcal{S}$}{
    	add $v$ to $\mathcal{S}$\;
    }{
    	add the other endpoint to $\mathcal{S}$\;
    }
  }
  remove all edges in $\mathcal{S}$\;
  re-calculate ${\mathbf r}$, $\frac{\partial f({\mathbf r})}{\partial {\mathbf A}}$ by Eq.~\eqref{eq:low_rank}, and $\frac{df({\mathbf r})}{d{\mathbf A}}$ by Eq.~\eqref{eq:gradient}\;
 }
 \Return $\mathcal{S}$\;
 \caption{\aurora-S}
 \label{alg:aurora_s}
\end{algorithm}

\begin{lm}
\label{lemma:complexity_subgraph}
  (Time and Space Complexities of \aurora-S). Algorithm 3 is $O(mk)$ in time and $O(m+n)$ in space, where $m$ and $n$ are the numbers of edges and nodes in the input graph; and $k$ is the budget.
\end{lm}
\begin{IEEEproof}
\hide{ 
It takes $O(m)$ time complexity to calculate ${\mathbf r}$ and $\frac{\partial f({\mathbf r})}{\partial {\mathbf A}}$ by applying power iterations. In the while-loop, we find the edge with the greatest influence by traversing all edges, which takes $O(m)$ time. Time spent to re-calculate ${\mathbf r}$ and $\frac{\partial f({\mathbf r})}{\partial {\mathbf A}}$ remains the same as $O(m)$. Since the body inside loop will run $k$ times, the overall time complexity is $O(mk)$. In Algorithm 3, it takes $O(m)$ space to save the sparse adjacency matrix ${\mathbf A}$ and $O(n)$ space to save the PageRank vector ${\mathbf r}$ and column vector $\mathbf{Q}'\mathbf{r}$ in Eq.~\eqref{eq:low_rank}. Therefore it has $O(m+n)$ space complexity.
}
Omitted for space.
\end{IEEEproof}

\subsection{Generalization and Variants}
The proposed family of \aurora\ algorithms assume the input graph is a plain network. However, it is worth pointing out that \aurora\ algorithms also work on different types of networks and other random walk-based techniques.

\paragraph{\aurora\ on Normalized PageRank}
Recall that in Section~\ref{sec:prob}, we remove the constraint of ${\mathbf A}$ being a normalized adjacency matrix and use the general form instead. The effect of removing that constraint will cause the $L_1$ norm of PageRank to be not equal to 1. We show that \aurora\ is also able to work on $L_1$ normalized PageRank. Let $S({\mathbf r}) = \sum_i {\mathbf r}(i)$, we have
\begin{equation}
\label{eq:gradient_normalized}
\frac{\partial f(r)}{\partial \mathbf{A}} = c\mathbf{Q}'(-\frac{2f({\mathbf r})}{S({\mathbf r})}{\mathbf 1}+\frac{2}{S({\mathbf r})}{\mathbf r}){\mathbf r}'
\end{equation}
Then we can apply \aurora\ algorithms by replacing Eq.~\eqref{eq:low_rank} with Eq.~\eqref{eq:gradient_normalized}

\paragraph{\aurora\ on Network of Networks}
Network of Networks (NoN) is a type of networks first introduced in \cite{non} with the ability to leverage the within-network smoothness in the domain-specific network and the cross-network consistency through the main network. An NoN is usually defined as the triplet ${\mathcal R}=<{\mathbf G}, {\mathcal A}, \theta>$, where ${\mathbf G}$ is the main network, ${\mathcal A}$ is a set of domain-specific network and $\theta$ is a mapping function to map the main node to the corresponding domain-specific network. In \cite{non}, {\sc CrossRank} and {\sc CrossQuery} are two ranking algorithms proposed to solve ranking on NoN. The authors have proved that they are actually equivalent to the well-known PageRank and random walk with restart on the integrated graph. Thus, \aurora\ algorithms also have the ability to audit {\sc CrossRank} and {\sc CrossQuery} on Network of Networks.

\paragraph{\aurora\ on Attributed Networks}
Given a large attributed network, it is important to learn the most influential node-attribute or edge-attribute w.r.t. a query node. We show that \aurora\ algorithms have the ability to find top-$k$ influential edge-attributes and node-attributes on attributed networks. The central idea is to treat attributes as {\em attribute nodes} and form an augmented graph with those attribute nodes. To support node attributes, let ${\mathbf A}$ be the $a\times a$ node-to-node adjacency matrix, and ${\mathbf W}$ be the $w\times a$ node-to-attribute matrix, then we can form an augmented graph ${\mathbf G} = \bigl( \begin{smallmatrix}{\mathbf A} & {\mathbf 0}\\ {\mathbf W} & {\mathbf 0}\end{smallmatrix}\bigr)$. To support edge attributes, similar to \cite{pienta2014mage}, we embed an edge-node for each edge in the input graph and define a mapping function $\psi$ that maps each edge-node to edge attribute in the original graph. We assume ${\mathbf A}$ is the $a\times a$ node-to-node adjacency matrix, and $b$ is the number of different edge attribute values. By embedding edge-node, it creates a $(a+b)\times (a+b)$ augmented graph ${\mathbf Y}$. Then to find the top-$k$ influential node-attributes and edge-attributes, we can easily run \aurora-N on the augmented attributed graphs ${\mathbf G}$ and ${\mathbf Y}$, respectively.

\section{Experimental Evaluation}\label{sec:exp}
In this section, we evaluate the proposed \aurora\ algorithms. All experiments are designed to answer the following two questions:
\begin{itemize}
\item{\textbf{Effectiveness.}} How effective are the proposed \aurora\ algorithms in identifying key graph elements w.r.t. the PageRank results?
\item{\textbf{Efficiency.}} How efficient and scalable are the proposed \aurora\ algorithms?
\end{itemize}

\subsection{Setup}\label{sec:setup}

\textbf{Datasets.} We test our algorithms on a diverse set of real-world network datasets. All datasets are publicly available. The statistics of these datasets are listed in Table~\ref{tab:datasets}.

\begin{table}
  \centering
  \caption{Statistics of the datasets}
  \label{tab:datasets}
  \begin{tabular}{c|c|c|c|c}
    \toprule
    {\bf Category} & {\bf Network} & {\bf Type} & {\bf Nodes} & {\bf Edges} \\
    \midrule
    \multirow{4}{*}{\textsc{Social}}
    & Karate & U & 34 & 78 \\
    & Dolphins & U & 62 & 159 \\
    & WikiVote & D & 7,115 & 103,689 \\
    & Pokec & D & 1,632,803 & 30,622,564 \\
    \hline
    \multirow{6}{*}{\textsc{Collaboration}}
    & GrQc & U & 5,242 & 14,496 \\
    & DBLP & U & 42,252 & 420,640 \\
    & NBA & U & 3,923 & 127,034 \\
    & cit-DBLP & D & 12,591 & 49,743 \\
    & cit-HepTh & D & 27,770 & 352,807 \\
    & cit-HepPh & D & 34,546 & 421,578 \\
    \hline
    \multirow{1}{*}{\textsc{Physical}}
    & Airport & D & 1,128 & 18,736 \\
    \hline
    \multirow{2}{*}{\textsc{Others}}
    & Lesmis & U & 77 & 254 \\
    & Amazon & D & 262,111 & 1,234,877 \\
    \hline
    \bottomrule
    \hline
    \multicolumn{5}{c}{(In Type, U means undirected graph; D means directed graph.)}
  \end{tabular}
  \vspace{-6mm}
\end{table}
\begin{itemize}
  \item \textsc{Social Networks}. Here, nodes are users and edges indicate social relationships. Among them,
  \textit{Karate} \cite{zachary1977information} is a well-known network dataset of a university karate club collected by Wayne Zachary in 1977.
  \textit{Dolphins} \cite{lusseau2003bottlenose} is an undirected social network of frequent associations between dolphins in a community living off Doubtful Sound, New Zealand.
  \textit{WikiVote}~\cite{leskovec2010signed} is generated by Wikipedia voting data from the inception of Wikipedia till January 2008.
  \textit{Pokec}~\cite{takac2012data} is a popular online social network in Slovakia.
  \item \textsc{Collaboration Networks}. Here, nodes are individuals and two people are connected if they have collaborated together.
  We use the collaboration network in the field of General Relativity and Quantum Cosmology (\textit{GrQc}) in Physics from arXiv preprint archive\footnote{https://arxiv.org/}.
  \textit{DBLP}\footnote{http://dblp.uni-trier.de/} is a co-authorship network from DBLP computer science bibliography.
  And \textit{NBA}~\cite{li2015replacing} is a collaboration network of NBA players from 1946 to 2009.
  \textit{cit-DBLP}~\cite{ley2002} is the citation network of DBLP, a database of scientific publications such as papers and books. Each node in the network is a publication, and each edge represents a citation of a publication by another publication.
  \textit{cit-HepTh}~\cite{leskovec2005graphs} is an ArXiv HEP-TH (High Energy Physics - Theory) citation network. The data covers papers from January 1993 to April 2003. If a paper $i$ cites paper $j$, there is a directed edge from $i$ to $j$.
  \textit{cit-HepPh}~\cite{leskovec2005graphs} is an ArXiv HEP-PH (High Energy Physics - Phenomenology) citation network. The data covers papers from January 1993 to April 2003. If a paper $i$ cites paper $j$, there is a directed edge from $i$ to $j$.
  \item \textsc{Physical Infrastructure Networks}. This category refers to the networks of physical infrastructure entities. Nodes in them correspond to physical infrastructure, and edges are connections.
  \textit{Airport}\footnote{https://www.transtats.bts.gov/} is a dataset of airline traffic. Each node represents an airport in the United States, an edge $(i, j)$ represents the airline from $i$ to $j$ while the edge weight stands for the normalized number of passengers.
  \item \textsc{Others}. This category contains networks that do not fit into the above categories.
  \textit{Lesmis} \cite{knuth1993stanford} is a network of co-appearances of characters in Victor Hugo's novel \textit{"Les Miserables"}. A node represents a character and an edge connects a pair of characters if they both appear in the same chapter of the book.
  \textit{Amazon} \cite{leskovec2007graph} is a co-purchasing network collected by crawling Amazon website. It is based on the {\it Customers Who Bought This Item Also Bought} feature.
\end{itemize}

\textbf{Baseline Methods.} We compare our proposed methods with several baseline methods, which are summarized as follows.
\begin{itemize}
  \item {\it Brute force (brute force)}. Calculate the changes by iterating all possible combinations of graph elements.

  \item {\it Random Selection (random)}. Randomly select $k$ elements and calculate the change by removing them.

  \item {\it Top-$k$ Degrees (degree)}. We first define the degree of an edge $(u, v)$ as follows,
  \begin{displaymath}
    d(u, v) =
    \begin{cases}
      (d(u) + d(v))\times \underset{i\in\{u, v\}}\max d(i), & if\ undirected \\
      (d(u) + d(v))\times d(u), & if\ directed
    \end{cases}
  \end{displaymath}
where $d(u)$ represents the degree of node $u$.

To audit by graph elements, we select $k$ elements with the highest degrees. For edges, we select $k$ edges with the highest edge degrees defined above; for nodes, we  select $k$ nodes with the highest node degrees; for subgraphs, we form a vertex-induced subgraph from $k$ nodes with the highest degrees.

  \item {\it PageRank}. We first define the PageRank score of an edge $(u, v)$ as follows,
  \begin{displaymath}
    {\mathbf r}(u, v) =
    \begin{cases}
      ({\mathbf r}(u) + {\mathbf r}(v))\times \underset{i\in\{u, v\}}\max {\mathbf r}(i), & if\ undirected \\
      ({\mathbf r}(u) + {\mathbf r}(v))\times {\mathbf r}(u), & if\ directed
    \end{cases}
  \end{displaymath}
where $r(u)$ is the PageRank score of node $u$.

To audit by graph elements, we select $k$ elements with the highest PageRank scores. That is, for edges, we select $k$ edges with the highest PageRank scores defined above; for nodes, we select $k$ nodes with highest PageRank scores; for subgraphs, we form a vertex-induced subgraph from $k$ nodes with the highest PageRank scores.

  \item {\it HITS}. We first define HITS score of an edge $(u, v)$ and node $u$ as follows,
  \vspace{-2mm}
  \begin{displaymath}
  \vspace{-2mm}
    HITS(u, v) = hub(u) \times hub(v) + auth(u) \times auth(v)
  \end{displaymath}
  \begin{displaymath}
  \vspace{-2mm}
    HITS(u) = hub(u) + auth(u)
  \end{displaymath}
where $hub(u)$ and $auth(u)$ represent the hub score and authority score of node $u$, respectively.

To audit by graph elements, we select $k$ elements with the highest HITS scores. That is, for edges, we select $k$ edges with the highest HITS scores defined above; for nodes, we select $k$ nodes with the highest HITS scores; for subgraphs, we form a vertex-induced subgraph from $k$ nodes with the highest HITS scores.
\end{itemize}

\textbf{Metrics}. Here, we choose the loss function to be squared $L_{2}$ norm. We quantify the performance of auditing by the goodness score $\Delta{f}$ (defined in Eq.~\eqref{eq:optimization}) of the graph elements $\mathcal{S}$ found by the corresponding algorithms.  

\textbf{Repeatability and Machine Configuration}. All datasets are publicly available. We will release the code of our proposed algorithms upon the publication of the paper. All experiments are performed on a virtual machine with 4 Intel i7-8700 CPU cores at 3.4GHz and 32GB RAM. The operating system is Windows 10. All codes are written in Python 3.6.

\begin{figure*}
\centering
\includegraphics[width=\textwidth]{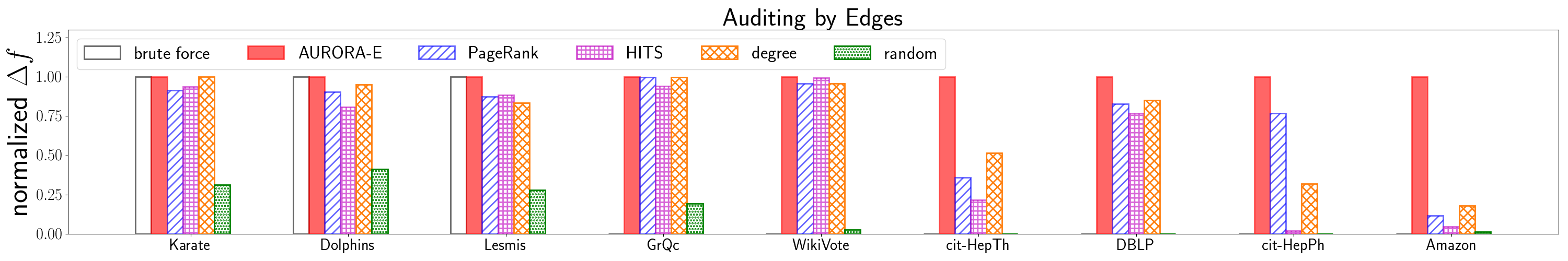}
\vspace{-6mm}
\caption{Auditing results by edges. Budget $k=10$. Higher is better. Best viewed in color.}
\vspace{-4mm}
\label{fig:edge_effective}
\end{figure*}

\begin{figure*}
\centering
\includegraphics[width=\textwidth]{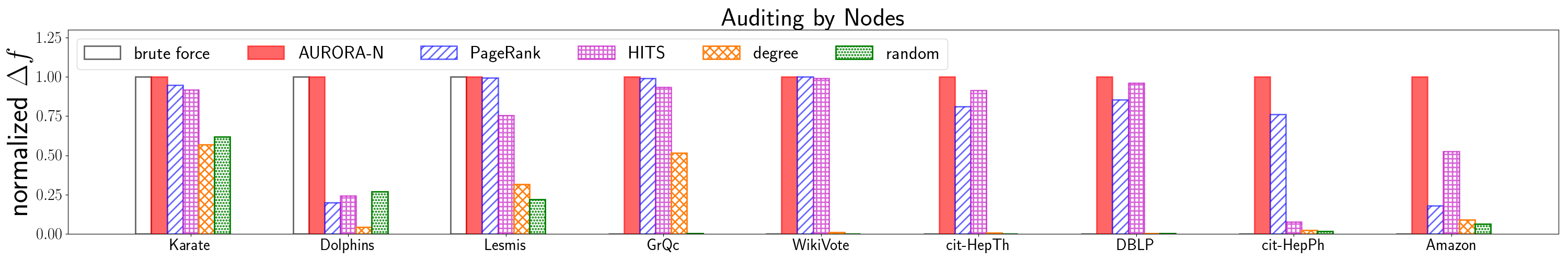}
\vspace{-6mm}
\caption{Auditing results by nodes. Budget $k=10$. Higher is better. Best viewed in color.}
\vspace{-4mm}
\label{fig:node_effective}
\end{figure*}

\begin{figure*}
\centering
\includegraphics[width=\textwidth]{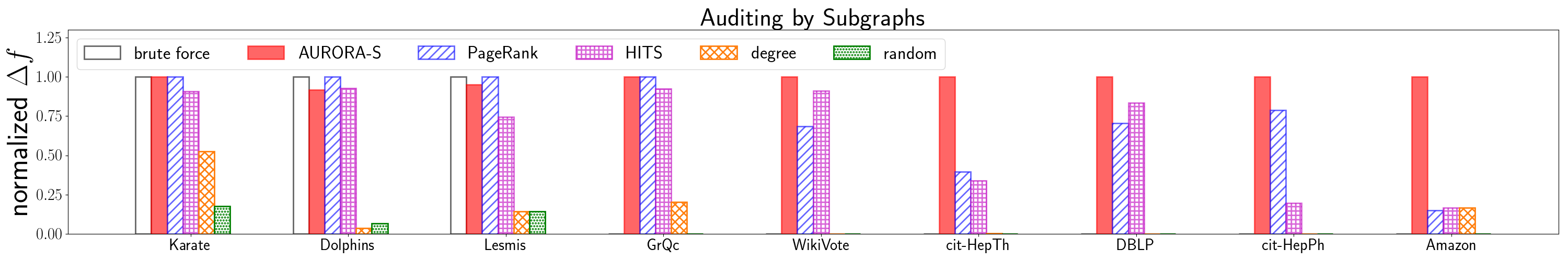}
\vspace{-6mm}
\caption{Auditing results by subgraphs. Budget $k=10$. Higher is better. Best viewed in color.}
\vspace{-4mm}
\label{fig:subgraph_effective}
\end{figure*}

\begin{figure*}[h]
  \begin{subfigure}{0.33\textwidth}
  \includegraphics[width=\textwidth]{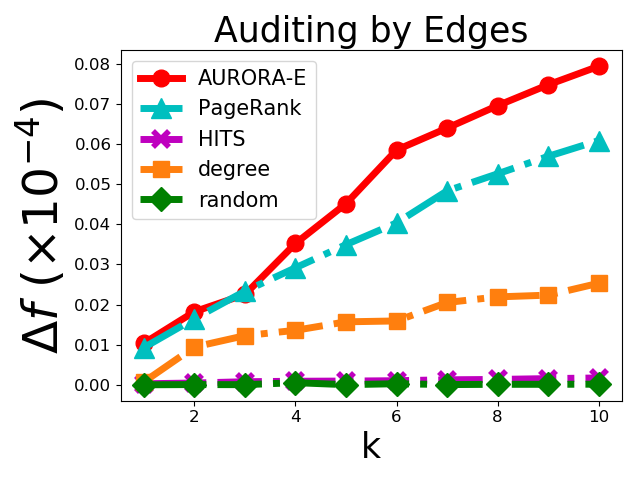}
  \vspace{-8mm}
  \caption{Auditing by Edges}
  \label{fig:edge_different_k}
  \end{subfigure}
  \begin{subfigure}{0.33\textwidth}
    \includegraphics[width=\textwidth]{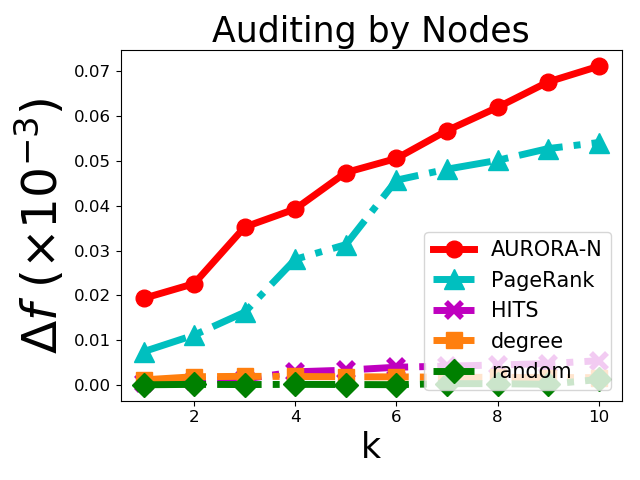}
    \vspace{-8mm}
    \caption{Auditing by Nodes}
    \label{fig:node_different_k}
  \end{subfigure}
  \begin{subfigure}{0.33\textwidth}
    \includegraphics[width=\textwidth]{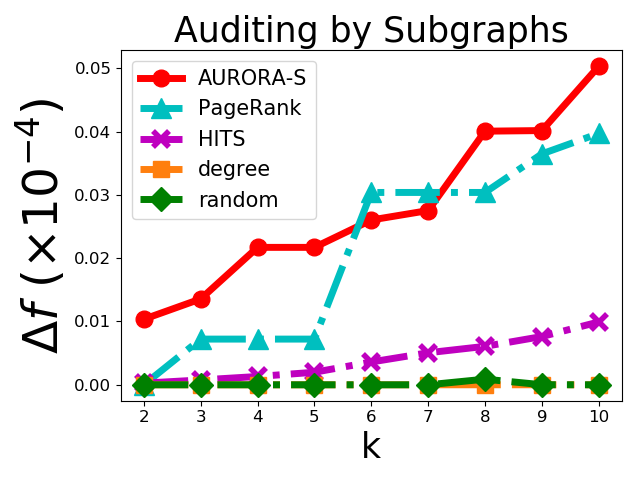}
    \vspace{-8mm}
    \caption{Auditing by Subgraphs}
    \label{fig:subgraph_different_k}
  \end{subfigure}
  \vspace{-1mm}
  \caption{Effect of $k$ on auditing results (cit-HepPh Dataset). Higher is better. Best viewed in color.}
  \vspace{-6mm}
  \label{fig:cit-hepph}
\end{figure*}

\begin{figure*}[h]
  \begin{subfigure}{0.33\textwidth}
  \includegraphics[width=\textwidth]{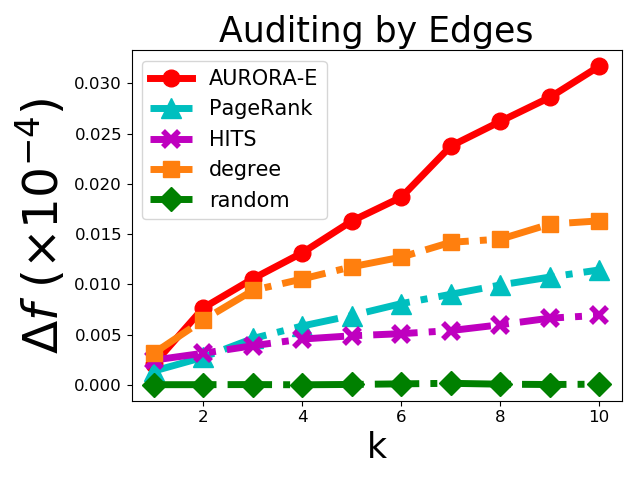}
  \vspace{-8mm}
  \caption{Auditing by Edges}
  \label{fig:edge_different_k}
  \end{subfigure}
  \begin{subfigure}{0.33\textwidth}
    \includegraphics[width=\textwidth]{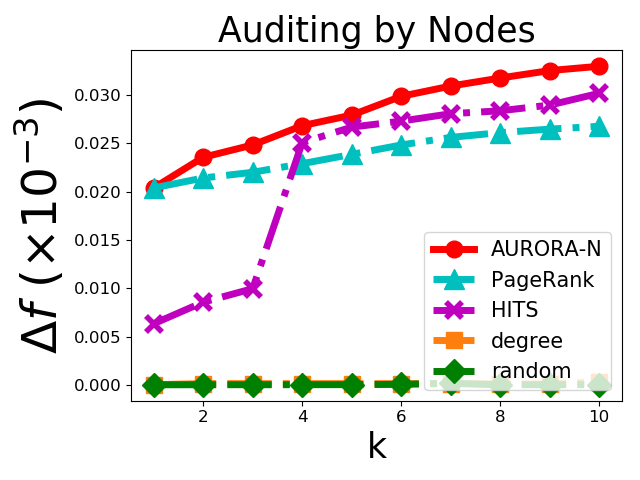}
    \vspace{-8mm}
    \caption{Auditing by Nodes}
    \label{fig:node_different_k}
  \end{subfigure}
  \begin{subfigure}{0.33\textwidth}
    \includegraphics[width=\textwidth]{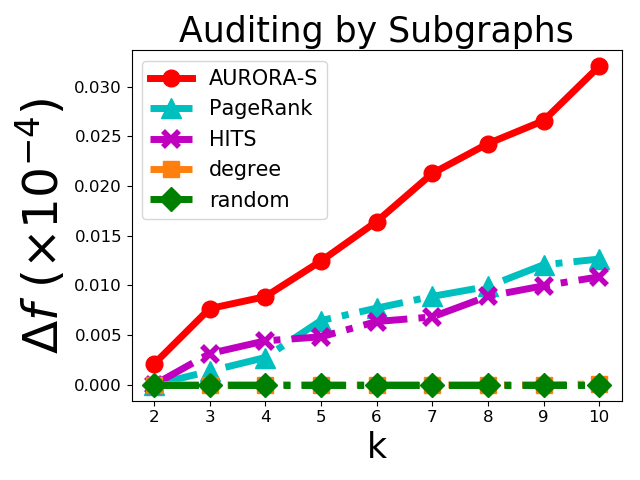}
    \vspace{-8mm}
    \caption{Auditing by Subgraphs}
    \label{fig:subgraph_different_k}
  \end{subfigure}
  \caption{Effect of $k$ on auditing results (cit-HepTh Dataset). Higher is better. Best viewed in color.}
  \vspace{-6mm}
  \label{fig:cit-hepth}
\end{figure*}

\subsection{Effectiveness Results}
\paragraph{\textbf{Quantitative Comparison}}
We perform effectiveness experiments with the baseline methods. We set $k$ from $1$ to $10$ and find $k$ influential edges and nodes, respectively. We set $k$ only from $2$ to $10$ to find an influential subgraph of size-{\em k}, respectively. This is because a vertex-induced subgraph with only $1$ node does not contain any edge and therefore is meaningless for the purpose of PageRank auditing. We conduct the experiments with \textit{brute force} on relatively small datasets to obtain the ground-truth of Eq.~\eqref{eq:optimization}, including \textit{Karate}, \textit{Dolphins} and \textit{Lesmis}. On other larger datasets, searching a ground-truth with $k$ most influential elements is prohibitively expensive due to its combinatorial nature. For example, even if we use the small \textit{Lesmis} dataset, it will take over a day to find ground-truth with $k=5$. Therefore, the results by {\em brute force} are absent on the remaining larger datasets.

The results of quantitative comparison across 9 different datasets are shown from Figure~\ref{fig:edge_effective} to Figure~\ref{fig:subgraph_effective}. From those figures, we have the following observations: (1) our family of \aurora\ algorithms consistently outperform other baseline methods (except for brute force) on all datasets; (2) on the three small datasets with ground-truth {\em brute force}, the performance of the proposed \aurora\ algorithms is very close to the ground-truth. Figure~\ref{fig:cit-hepph} and Figure~\ref{fig:cit-hepth} shows the effect of k on auditing results. We can observe the following findings from those figures: (1) our family of \aurora\ algorithms incrementally find influential graph elements w.r.t. the budget $k$; (2) the proposed \aurora\ algorithms consistently outperform baseline methods on different budgets.

\paragraph{\textbf{Case Studies on {\em Airport} Dataset}} A natural use case of our \aurora\ algorithms is to find influential edges and nodes in a given graph. To demonstrate that our algorithms are indeed able to provide intuitive information, we test our algorithms on the \textit{Airport} dataset. This dataset was manually created from commercial airline traffic data in 2017, which is provided by United States Department of Transportation. More detailed description and statistics of this dataset can be found in Subsection~\ref{sec:setup}. We perform \aurora-E and \aurora-N to find the most influential airlines (edges in the graph) and airports (nodes in the graph) across United States with $k=7$.

Edges selected by \aurora-E are {\tt ATL-LAX}, {\tt LAX-ATL}, {\tt ATL-ORD}, {\tt ORD-ATL}, {\tt ATL-DEN}, {\tt DEN-ATL} and {\tt LAX-ORD}. In contrast, PageRank selects {\em ATL-LAS} instead of {\tt DEN-ATL} and {\em ATL-DFW} instead of {\tt LAX-ORD}. {\tt DEN-ATL} plays a more important role in determining the centrality (e.g., PageRank) of other airports. This is because {\tt DEN} serves as one of the busiest hub airports that connects West coast and East coast; while {\em ATL-LAS} is less important in that regard, considering the existence of {\tt ATL-LAX} and {\tt ATL-PHX}. Comparing {\tt LAX-ORD} and {\em ATL-DFW}, {\tt LAX-ORD} directly connects Los Angeles and Chicago, both of them are largest cities in the United States.

In the scenario of node-auditing, \aurora-N selects {\tt ATL}, {\tt LAX}, {\tt ORD}, {\tt DFW}, {\tt DEN}, {\tt LAS} and {\tt CLT}. In contrast, PageRank selects {\em SFO} instead of {\tt CLT}.  {\tt CLT} seems to be a more reasonable choice because it serves as a major hub airport, the $6^{\textrm{th}}$ busiest airport by FAA statistics, to connect many regional airports around States like North Carolina, South Carolina, Virginia, West Virginia, etc. Compared with {\tt CLT}, {\em SFO} is less influential in that regard, mainly due to the following two reasons: (1) it ranks after {\tt CLT} ( $7^{\textrm{th}}$ vs. $6^{\textrm{th}}$) in the list of busiest airports by FAA statistics; (2) due to the location proximity of {\em SFO} to {\tt LAX} and {\tt SJC}, even if this node is perturbed (i.e., absent), many surrounding airports (especially regional airports in California) could still be connected via {\tt LAX} and {\tt SJC}.

\paragraph{\textbf{Case Studies on {\em DBLP} Dataset}} Another interesting use case of \aurora\ algorithms is sense-making in graph proximity. We construct a co-authorship network from DBLP computer science bibliography to test our algorithms. We perform \aurora-N\ and PageRank with $k=6$. Different from the previous case study, here we use a personalized PageRank with the query node is {\tt Christos Faloutsos}. In this case, the top-ranked scholars in the resulting ranking vector $\mathbf r$ form the proximity (i.e., `neighborhood') of the query node (i.e., who are most relevant to {\tt Christos Faloutsos}). Consequently, the nodes selected by an auditing algorithm indicate those important nodes in terms of making/maintaining the neighborhood of the query node. Comparing the results by \aurora-N and PageRank, 6 of them are the same while \aurora-N selects {\tt Jure Leskovec} instead of {\em Yannis Ioannidis}. This result is consistent with the intuition, since {\tt Jure Leskovec}, as the former student of {\tt Christos Faloutsos} with lots of joint publications, plays a more prominent role in the neighborhood of {\tt Christos Faloutsos} by sharing more common collaborators. 

\paragraph{\textbf{Case Studies on {\em NBA} Dataset}} In a collaboration network, a subgraph can be naturally viewed as a team (e.g. sports team). From this perspective,  \aurora-S has the potential to find teammates of a player. We set the query node as {\tt Allen Iverson}. Since there are $5$ players for each team on the court, we set $k=5$. The subgraph selected by \aurora-S consists of {\tt Allen Iverson}, {\tt Larry Hughes}, {\tt Theo Ratliff}, {\tt Joe Smith} and {\tt Tim Thomas}. In contrast, in the resulting subgraph by PageRank, {\tt Tim Thomas} is replaced by {\em Drew Gooden}. Though {\em Drew Gooden} and {\tt Allen Iverson} shares many common neighbors in the network, which might cause its ranking in PageRank to be higher, they have never played in the same team together. Compared with the subgraph by PageRank, all players in the subgraph by \aurora-S have played with {\tt Iverson} in Philadelphia Sixers during the time period from 1997 to 1999. 

\begin{figure}
  \centering
  \minipage[t]{0.22\textwidth}
    \centering
    \includegraphics[width=\textwidth]{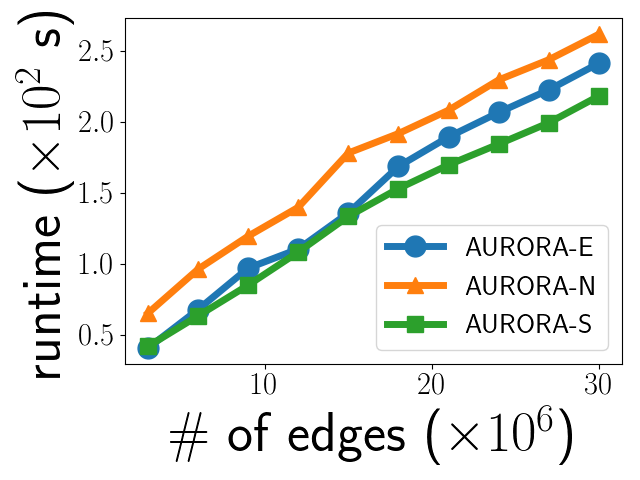}
    \caption{running time vs. the number of edges on Pokec dataset}
    \label{fig:scalablity_edges}
  \endminipage\hfill
  \minipage[t]{0.22\textwidth}
    \centering
    \includegraphics[width=\textwidth]{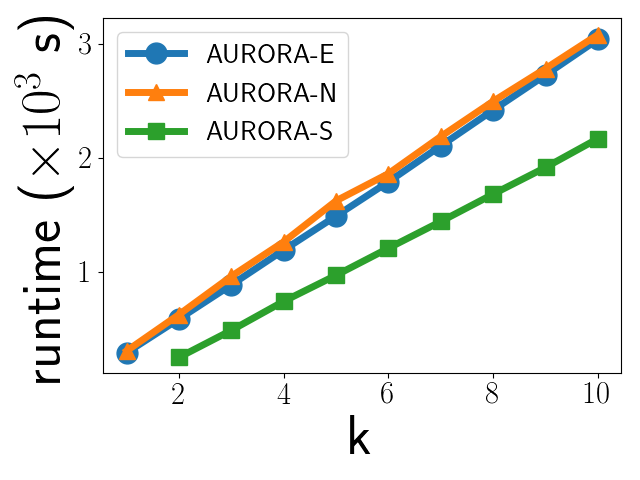}
    \caption{running time vs. number of $k$ on Pokec dataset}
    \label{fig:scalability_k}
  \endminipage\hfill
  \vspace{-6mm}
\end{figure}
\subsection{Efficiency Results}
We show the running time vs. number of edges $m$ and budget size $k$ on \textit{Pokec} dataset in Figure~\ref{fig:scalablity_edges} and Figure~\ref{fig:scalability_k}. We can see that the proposed \aurora\ algorithms scale linearly with respect to $m$ and $k$, respectively. This is consistent with our complexity analysis that the family of \aurora\ algorithms are linear with respect to the number of edges and the budget.

\begin{figure}
  \centering
    \includegraphics[width=0.48\textwidth]{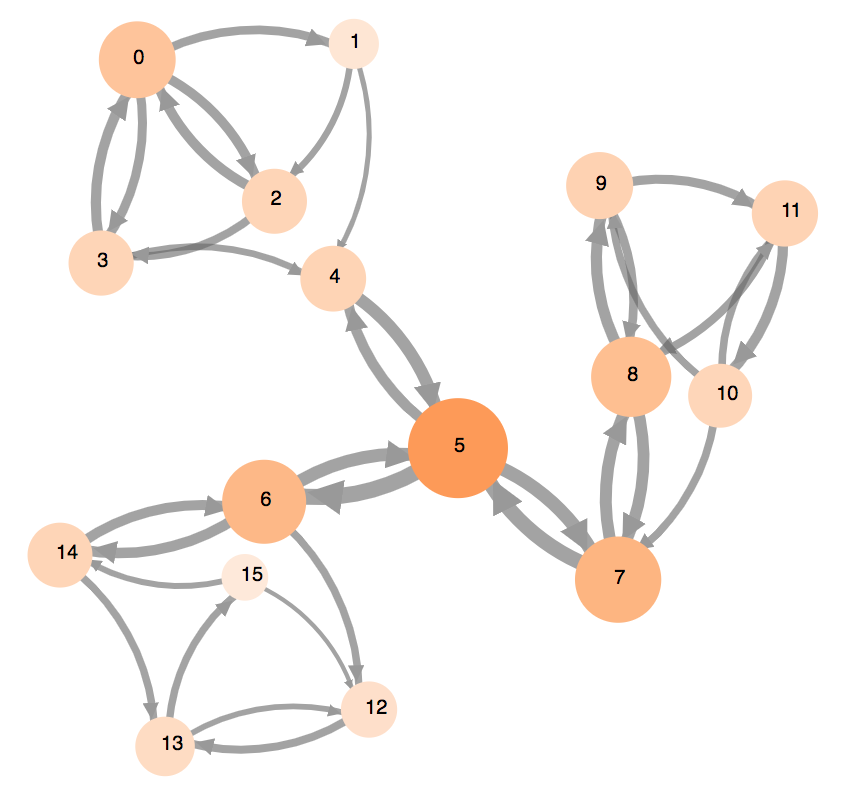}
    \caption{Visualization of toy graph on the visualization system. Best viewed in color.}
    \label{fig:exp_vis}
    \vspace{-8mm}
\end{figure}

\subsection{Visualization}
To better understand the auditing results, we developed a prototype system with D3.js to represent the influence of graph elements visually. In the system, we use the strength of line to represent the gradient of an edge and use the size and color for the gradient of nodes. An example of visualizing hand-crafted toy graph is shown in Figure~\ref{fig:exp_vis}. It is easy to see in the figure that Node $5$ is the most influential node, and edges around Node $5$ is more influential than other edges, both of which are consistent with our intuition.

\section{Related Work}\label{sec:related}
In this section, we briefly review the related work, from the following two perspectives, including (1) graph ranking and (2) explainability.

\textbf{Graph ranking}. Regarding graph ranking, PageRank \cite{ilprints422} and HITS \cite{kleinberg1999} are probably the most well-known and widely used algorithms. PageRank measures the importance of nodes as a stationary distribution of random walks. HITS assumes that each node has two scores: \textit{hub} and \textit{authority}. A node has a high hub score if it links to many nodes with high authority scores, and a node has a high authority score if it links to many nodes with high hub scores. Many variants of PageRank and HITS have been developed in the literature. To name a few, in \cite{ng2001}, the author studies the stability of PageRank and HITS, based on which they propose two new algorithms (Randomized HITS and Subspace HITS). Ding et al. \cite{ding2003} provide a unified ranking method for HITS and PageRank. In \cite{haveliwala2002topic}, Haveliwala et al. propose the well known \textit{personalized PageRank} by replacing the uniform teleportation vector with a biased personalized topic-specific vector; while  random walk with restart \cite{tong2006fast} concentrates all teleportation probabilities to a single node. Other random walk based graph ranking methods include \cite{jeh2002simrank} and many more. 

\textbf{Explainability}. How to enhance interpretability of machine learning and data mining models has been attracting a lot of research interests in recent years. 
Two representative ways to explain such black-box predictions consist of (a) using other interpretable models to provide interpretable representation of model's predictions~\cite{caruana2015, ribeiro2016, bastani2017}, and (b) quantifying the influence by perturbing either features or training data~\cite{feldman2015, adler2016, datta2016, li2016, pang2017}. The proposed \aurora\ algorithms follow the generic principle in \cite{pang2017} in the context of graph ranking, and has been integrated into a prototype system \cite{kang2018xrank}. Finding influential nodes on graphs is a very active research area. The central theme of these research is to find nodes to maximize the spread of influence in social networks (i.e., influence maximization)~\cite{gruhl2004information, chen2009efficient, gomez2010inferring}. The seminal work in influence maximization problem is attributed to 
 Kempe et al.~\cite{kempe2003,kempe2005}, in which they have discovered its diminishing returns property. The focus of this paper is fundamentally different from the classic influence maximization problem, in the sense that we want to find most influential nodes, edges or subgraphs w.r.t. the graph ranking results, as opposed to the size of `infected' nodes during the influence propagation process. 

\section{Conclusion}\label{sec:conclusion}
In this paper, we study the problem of auditing PageRank, where we aim to find the most influential graph elements (e.g., edges, nodes, subgraphs) w.r.t. graph ranking results. We formally define the PageRank auditing problem by measuring influence of each graph element as the rate of change in a certain loss function defined over the ranking vector, and formulate it as an optimization problem. We further propose a family of fast approximation algorithms, named \aurora, with $(1-1/e)$ approximation ratio and a linear complexity in both time and space. The extensive experimental evaluations on more than 10 datasets demonstrate that the proposed \aurora\ algorithms are able to identify influential graph elements, consistently outperform baseline methods, and scale linearly to large graphs. In the future, we would like to generalize this auditing paradigm to other graph ranking methods (e.g., HITS) as well as dynamic networks.

\section*{Acknowledgment}
This work is supported by NSF (IIS-1651203, IIS-1715385 and IIS-1743040), DTRA (HDTRA1-16-0017), ARO (W911NF-16-1-0168), DHS (2017-ST-061-QA0001), NSFC (61602306, Fundamental Research Funds for the Central Universities), and gifts from Huawei and Baidu.

\balance
\bibliographystyle{IEEEtran}
\bibliography{ref}

\end{document}